\documentclass[11pt,twoside]{article}
\usepackage{asp2010}

\resetcounters

\begin{document}

\title{The pulsating low-mass He-core white dwarfs} 

\author{A. H. C\'orsico$^{1,2}$, 
        L. G. Althaus$^{1,2}$, and 
        A. D. Romero$^3$
\affil{$^1$Facultad de Ciencias Astron\'omicas y Geof\'isicas, 
           Universidad Nacional de La Plata, 
           Paseo del Bosque s/n, 
           (1900) La Plata, Argentina}
\affil{$^2$Instituto de Astrof\'{\i}sica La Plata, 
                CONICET-UNLP, Argentina}
\affil{$^3$Departamento de Astronomia,
                Universidade Federal do Rio Grande do Sul,\\
                Av. Bento Goncalves 9500, Porto Alegre 91501-970, RS, 
                Brazil}}

\begin{abstract}
Recent years have witnessed  the discovery of many low-mass ($\lesssim
0.45  M_{\odot}$) white  dwarf (WD)  stars ---  expected to  harbor He
cores---  in  the field  of  the Milky  Way  and  in several  galactic
globular and open clusters.  Recently, three pulsating objects of this
kind have been discovered:  SDSS J1840+6423, SDSS J1112+1117, and SDSS
J1518+0658.  Motivated by these very exciting findings, and in view of
the  valuable  asteroseismological  potential  of  these  objects,  we
present here the main outcomes  of a detailed theoretical study on the
seismic properties of low-mass He-core WDs based on fully evolutionary
models  representative  of these  objects.   This  study  is aimed  to
provide a theoretical basis from which to interpret present and future
observations of variable low-mass WDs.
\end{abstract}

\section{Introduction}

\begin{figure*} 
\begin{center}
\includegraphics[clip,width=10 cm]{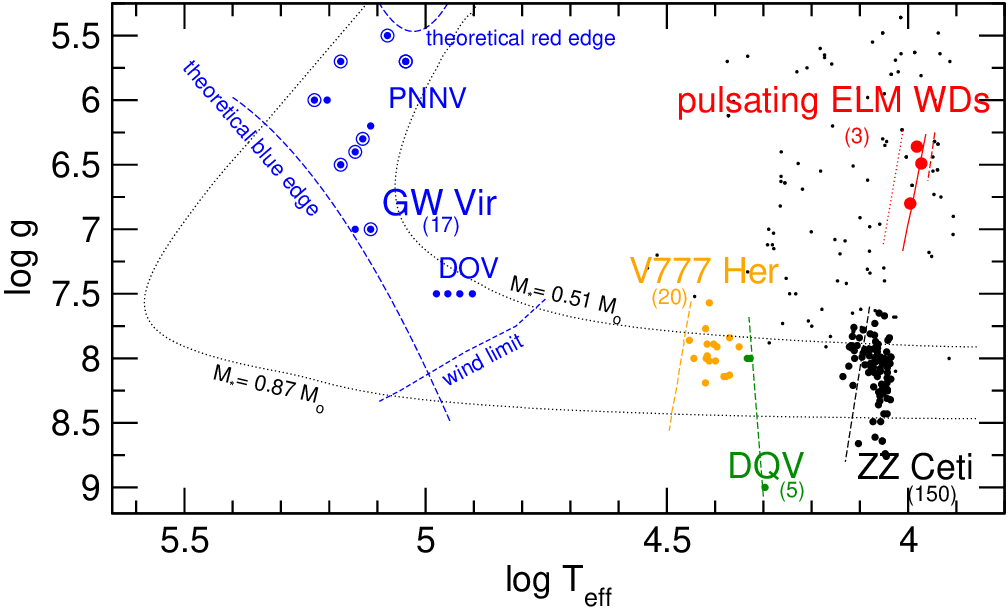} 
\caption{The location of the several  classes of pulsating WD stars in
  the $\log T_{\rm eff} - \log  g$ plane, marked with dots of differet
  colors. Two post-VLTP  (H-deficient) evolutionary tracks are plotted
  for  reference.  Also  shown is  the  theoretical blue  edge of  the
  instability strip for the GW  Vir stars (C\'orsico et al. 2006), the
  V777 Her stars  (C\'orsico et al.  2009a), the  DQV stars (C\'orsico
  et al. 2009b),  the ZZ Ceti stars (Fontaine  and Brassard 2008), and
  the pulsating  ELM WDs. For this  last class, we show  the blue edge
  according to  Hermes et al.  (2013) (dotted red line),  C\'orsico et
  al. (2012) (solid red line), and Steinfadt et al. (2010) (dashed red
  line). Small black dots correspond to low-mass WDs that either 
  are nonvariable or have not been observed to  vary.}
\label{figure1} 
\end{center}
\end{figure*}

The population of low-mass WDs  has masses lower than $0.45 M_{\odot}$
and    peaks    at    $\approx    0.39   M_{\odot}$    (Kleinman    et
al. 2013). Recently, a large  number of low-mass WDs with masses below
$\sim  0.20-0.25 M_{\odot}$ has  been discovered  (Brown et  al. 2010,
2012; Kilic  et al.   2011, 2012); they  are referred to  as extremely
low-mass (ELM)  WDs.  The low-mass WD population  is probably produced
by  strong mass-loss  episodes at  the  red giant  branch (RGB)  phase
before the He-flash  onset. As such, these WDs  are expected to harbor
He cores,  in contrast to average  mass WDs, which  all likely contain
C/O cores.  The internal structure of WDs can be disentangled by means
of asteroseismology (Winget \& Kepler 2008; Althaus et al. 2010), that
allows us to  place constraints on the stellar  mass, the thickness of
the  compositional layers,  and the  core chemical  composition, among
other  relevant properties.   Recently, three  pulsating ELM  WD stars
have been discovered: SDSS J184037.78+642312.3 ($T_{\rm eff}= 9390 \pm
140$ K and $\log g= 6.49 \pm 0.06$), SDSS J111215.82+111745.0 ($T_{\rm
  eff}=  9590 \pm  140$  K and  $\log  g= 6.36  \pm  0.06$), and  SDSS
J151826.68+065813.2 ($T_{\rm eff}=  9900 \pm 140$ K and  $\log g= 6.80
\pm 0.06$)  (Hermes et  al.  2012, 2013).   The exciting  discovery of
these stars  opens the possibility of applying  the powerful machinery
of  asteroseismology  to  these  low-mass, presumed  He-core  WDs.  In
Fig.  \ref{figure1} we  show a  $\log T_{\rm  eff} -  \log  g$ diagram
displaying the location of the different known classes of pulsating WD
stars: GW Vir  or pulsating PG1159 stars (blue dots),  V777 Her or DBV
stars  (orange dots), DQV  stars (green  dots), ZZ  Ceti or  DAV stars
(black dots),  and finally, the  newly discovered three  pulsating ELM
WDs (red dots).  According to the location of the pulsating ELM WDs in
this diagram,  they appear naturally as  the extension of  the ZZ Ceti
instability strip at the low gravity domain.

Motivated by  the asteroseismic  potential of pulsating  low-mass WDs,
and stimulated by the discovery of the first variable ELM WDs, we have
started  in  the La  Plata  Observatory  a  theoretical study  of  the
pulsation properties of low-mass, He-core WDs with masses in the range
$0.17-0.46 M_{\odot}$.  Here we present the main results of our study,
and  refer  to  the paper  by  C\'orsico  et  al.  (2012) for  a  full
description of the methods and findings.

\section{Adiabatiac properties}
\label{adiabatic}

We  have  analyzed  the   adiabatic  pulsation  properties  ($g$-  and
$p$-modes) of a  large set of low-mass and ELM WD  models extracted from the
computations  of  Althaus  et  al.   (2009).  Specifically,  we  have
considered 8 model  sequences of He-core WD generated  with the LPCODE
evolutionary code,  with stellar  masses $0.170, 0.198,  0.220, 0.251,
0.303, 0.358, 0.400$, and $0.452 M_{\odot}$, metallicity $Z= 0.03$ and
the  MLT for convection  with the  free parameter  $\alpha =  1.6$.  A
time-dependent  treatment of the  gravitational settling  and chemical
diffusion has been accounted for,  as well as residual nuclear burning
(see for details Althaus et  al.  2009). The pulsation computations of
our adiabatic survey were  performed with the nonradial pulsation code
described in detail  in C\'orsico \& Althaus (2006).  We have explored
the  adiabatic pulsation  properties  of these  models, including  the
expected  range  of  periods  and  period  spacings,  the  propagation
properties  and mode  trapping of  pulsations, the  regions  of period
formation, as well as the  dependence on the effective temperature and
stellar mass. For the first time, we assessed the pulsation properties
of  He-core   WDs  with  masses  in   the  range  $0.20-0.45
M_{\odot}$.  In particular,  we found  appreciable differences  in the
seismic properties  of objects with  $M_* \gtrsim 0.20  M_{\odot}$ and
the ELM WDs ($M_* \lesssim 0.20 M_{\odot}$). We summarize our
findings below:

\begin{itemize}
\item[-]  ELM WDs  have a  H envelope  that is  much thicker  than for
  massive He-core WDs ($\approx 0.20-0.45 M_{\odot}$), due to the very
  different  evolutionary   history  of  the   progenitor  stars.   In
  particular,  ELM  WDs   did  not  experience  diffusion-induced  CNO
  flashes, thus they harbor thick H envelopes.
\item[-] By virtue  of the thicker H envelope  characterizing ELM WDs,
  they experience H  burning via the pp chain,  and consequently their
  evolution  is  extremely  slow.   This  feature  makes  these  stars
  excellent candidates  to become pulsating  objects. However, He-core
  WDs with $M_* \approx  0.40-0.45 M_{\odot}$ should have evolutionary
  timescales of  the same order,  also making them  attractive targets
  for current searches of variable low-mass WDs.
\item[-] The thickness of the He/H transition region is markedly wider
  for the  ELM WDs than  for massive He-core  WDs. This is due  to the
  markedly lower surface gravity  characterizing ELM WDs, that results
  in  a less  impact of  gravitational settling,  and eventually  in a
  wider chemical transition.
\item[-]  The  Brunt-V\"ais\"al\"a  frequency  ($N$) for  ELM  WDs  is
  globally  lower than for  the case  of massive  objects, due  to the
  lower  gravity characterizing ELM  WDs. A  lower Brunt-V\"ais\"al\"a
  frequency profile leads to longer pulsation periods.
\item[-]  The  bump  of  $N^2$   at  the  He/H  transition  region  is
  notoriously more narrow and pronounced  for the massive WDs than for
  the ELM WDs.  This feature results in a weaker  mode trapping in ELM
  WDs, something that severely limits their seismological potential to
  constrain the thickness of the H envelope.
\item[-]   As  already  noted   by  Steinfadt   et  al.   (2010),  the
  Brunt-V\"ais\"al\"a frequency of  the ELM WDs is larger  in the core
  than in  the envelope. This is  in contrast with the  case of models
  with $M_* \gtrsim 0.20  M_{\odot}$, in which the Brunt-V\"ais\"al\"a
  frequency  exhibits   larger  values  at  the   outer  layers,  thus
  resembling the situation encountered in ZZ Ceti stars. So, $g$-modes
  in  ELM WDs  probe  mainly the  stellar  core and  have an  enormous
  asteroseismic potential, as it  was first recognized by Steinfadt et
  al. (2010).
\item[-] Similarly  to ZZ Ceti  stars, the $g$-mode  asymptotic period
  spacing  (and the  periods themselves)  in low-mass  He-core  WDs is
  sensitive  primarily to  the stellar  mass, and  to a  somewhat less
  extent, to the effective temperature. Specifically, $\Delta \Pi^{\rm
    a}_{\ell}$  is longer  for lower  $M_*$ and  $T_{\rm  eff}$. Also,
  there is  a non-negligible  dependence with the  thickness of  the H
  envelope, where $\Delta \Pi^{\rm  a}_{\ell}$ is longer for thinner H
  envelopes  (small  $M_{\rm H}$).  Typically,  the asymptotic  period
  spacing range  from $\approx  55$ s for  $M_* = 0.45  M_{\odot}$ and
  $T_{\rm eff}  = 11\,500$ K,  up to $\approx  110$ s for $M_*  = 0.17
  M_{\odot}$ and $T_{\rm eff} = 8000$ K.
\end{itemize}

\subsection{Effects of element diffusion}

\label{diffusion}
\begin{figure} 
\begin{center}
\includegraphics[clip,width=10 cm]{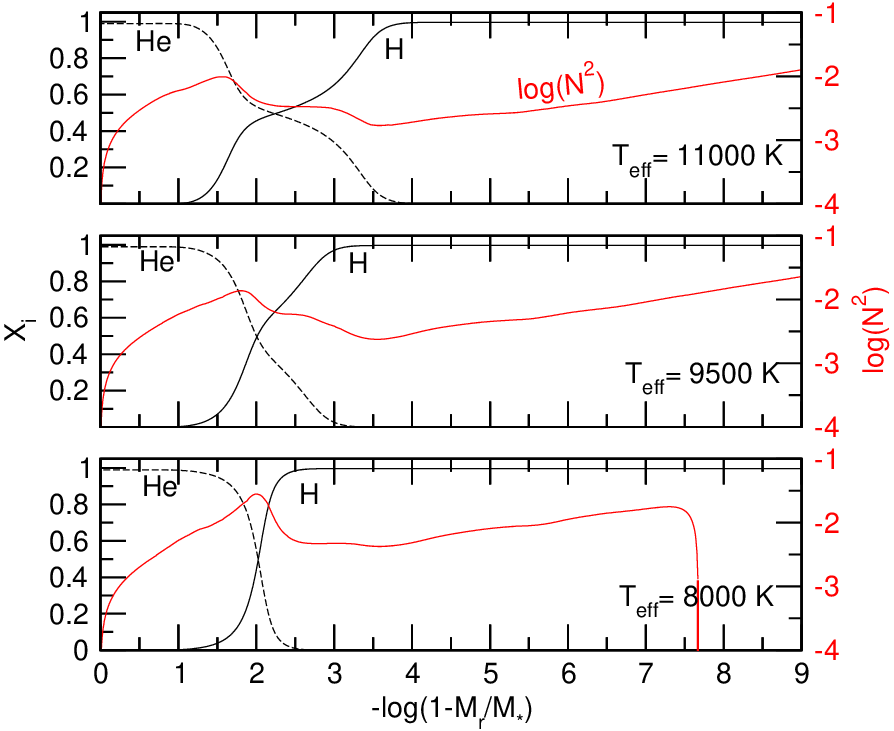} 
\caption{The internal chemical profile of  H and He, and the logarithm
  of the squared Brunt-V\"ais\"al\"a  frequency (red), for a ELM WD 
  model with $M_*
  =  0.17 M_{\odot}$  at different  effective temperatures,  which are
  indicated in each panel.}
\label{figure2} 
\end{center}
\end{figure}

Time-dependent element  diffusion modifies the  shape of the He  and H
chemical profiles as  the WD cools, causing H to  float to the surface
and He  to sink down. In  particular, diffusion not  only modifies the
chemical composition  of the outer layers,  but also the  shape of the
He/H  chemical transition  region  itself. This  is  clearly shown  in
Fig. \ref{figure2}  for the case  of the $0.17 M_{\odot}$  sequence in
the $T_{\rm  eff}$ interval ($11\, 000  - 8000$ K).  For  the model at
$T_{\rm  eff}  = 11\,000$  K,  the H  profile  is  characterized by  a
diffusion-modeled double layered chemical structure, which consists in
a pure H envelope atop an  intermediate remnant shell rich in H and He
(upper panel). This  structure still remains, although to  a much less
extent,   in  the   model  at   $T_{\rm   eff}  =   9500$  K   (middle
panel).  Finally, at  $T_{\rm eff}  =  8000$ K,  the H  profile has  a
single-layered  chemical structure  (lower  panel). Element  diffusion
processes affect  all the sequences considered in  this work, although
the transition from a double-layered structure to a single-layered one
occurs at different effective temperatures. The markedly lower surface
gravity that characterizes  the less massive model, results  in a less
impact of  gravitational settling, and eventually in  a wider chemical
transition in the  model with $M_* = 0.17  M_{\odot}$. Because of this
fact,  the  sequences  with  larger masses  reach  the  single-layered
structure at higher effective temperatures

The changes  in the  shape of the  He/H interface are  translated into
non-negligible  changes  in  the  profile of  the  Brunt-V\"ais\"al\"a
frequency,  as  can be  appreciated  in the  plot.  In  fact, at  high
effective  temperatures, $N^2$  is characterized  by two  bumps, which
merge  into  a  single one  when  the  chemical  profile at  the  He/H
interface  adopts a  single-layered  structure. In  the  light of  our
results,  we conclude  that  the assumption  diffusive equilibrium  is
valid in the  He/H transition region, which has  been adopted in other
works, is clearly wrong. We  found substantial changes in the value of
the  periods when diffusion  is neglected,  depending on  the specific
mode  considered and  the  value of  the  effective temperature.   For
instance, for the $k = 3$ mode ($\Pi_3 \approx 460$ s) at $T_{\rm eff}
\approx 9000$ K, a  variation of $5 \%$ in the value  of the period is
expected.

\section{Stability analysis}
\label{nonadiabatic}

\begin{figure} 
\begin{center}
\includegraphics[clip,width=10 cm]{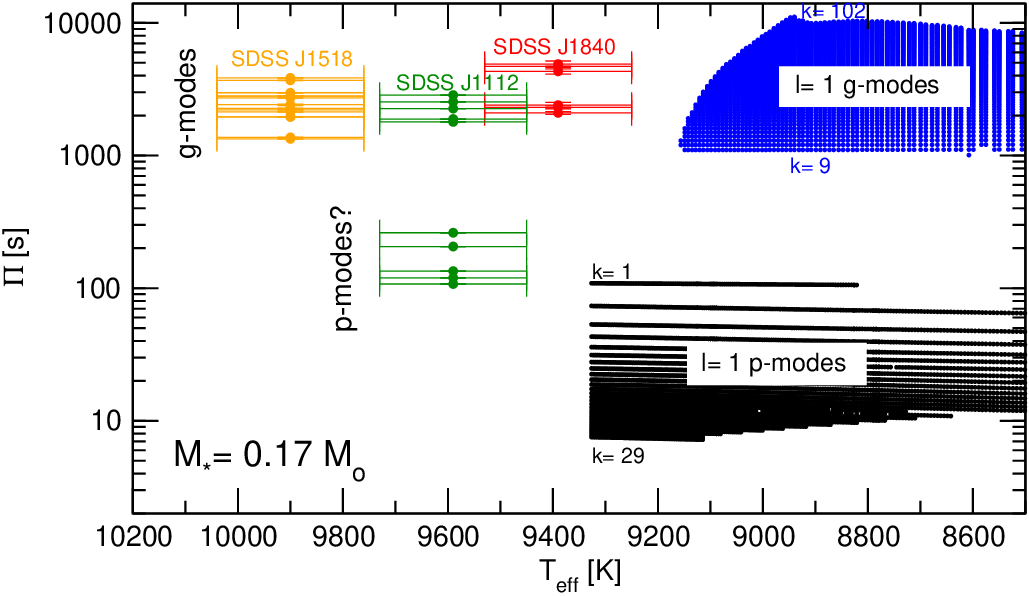} 
\caption{The instability  domain of  dipole $g$-modes (blue  dots) and
  $p$-modes   (black  dots)  on   the  $T_{\rm   eff}  -   \Pi$  plane
  corresponding to our set of models with $M_* = 0.17 M_{\odot}$. Also
  shown  are  the  periodicities  measured in  SDSS  J1840+6423,  SDSS
  J1112+1117,  and  SDSS  J1518+0658  (red, green,  and  orange  dots,
  respectively).}
\label{figure3} 
\end{center}
\end{figure}

\begin{figure} 
\begin{center}
\includegraphics[clip,width=10 cm]{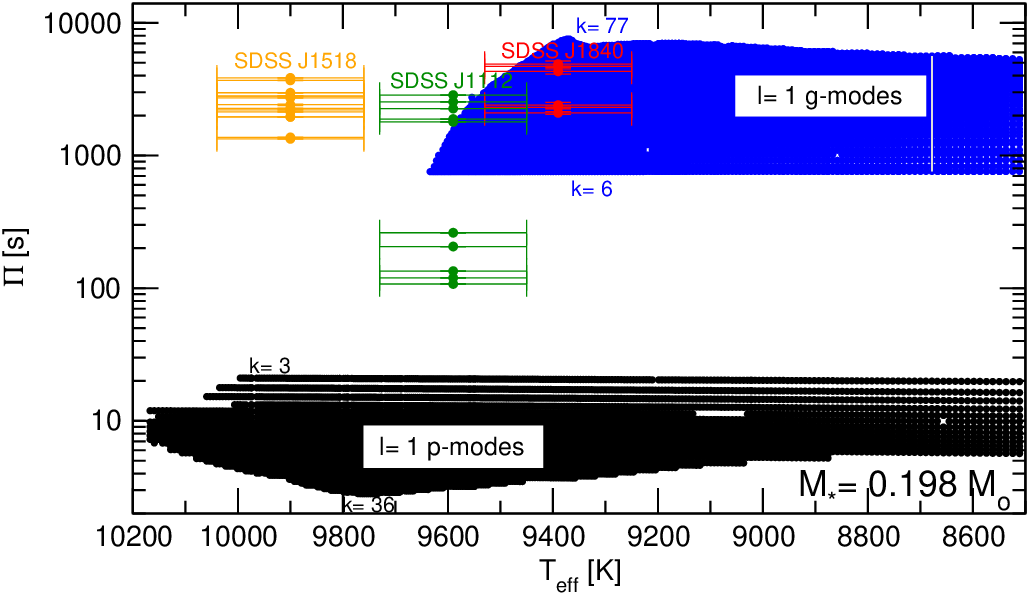} 
\caption{Same as in Fig. \ref{figure3}, but for our set of models with 
$M_* = 0.198 M_{\odot}$.}
\label{figure4} 
\end{center}
\end{figure}

In order  to investigate the plausibility of  excitation of pulsations
in  our  models, we  performed  a  linear  stability analysis  on  our
complete set  of evolutionary sequences. We  employed the nonadiabatic
pulsation  code described  in detail  in  Córsico et  al. (2006).  The
nonadiabatic computations rely on the frozen convection approximation,
in which the  perturbation of the convective flux  is neglected. While
this  approximation is  known  to give  unrealistic  locations of  the
$g$-mode red  edge of instability, it leads  to satisfactory predictions
for the  location of the  blue edge of  the ZZ Ceti  (DAV) instability
strip (van  Grootel et al. 2012).   We found that a  dense spectrum of
$g$-modes are excited by the ($\kappa - \gamma$)-mechanism acting in the
H  partial ionization  zone for  all the  masses considered,  and that
there exists a well-defined blue  (hot) edge of instability of He-core
WDs, which is the low-mass  analog to the blue edge of the ZZ
Ceti instability strip. We also found numerous unstable $p$-modes; the
corresponding blue edge is hotter  than the $g$-mode blue edge and has
a  lower slope.  We  warn that  the  location of  the  blue edges  are
sensitive to the efficiency  of convection adopted for the equilibrium
models. So, had we adopted a different convective efficiency, then the
predicted  blue edges  of the  ELM  and low-mass  He-core WDs
instability domain should appreciably change. In Fig. \ref{figure1} we
show the  location of our $g$-mode  blue edge (solid  red line), along
with those obtained  by Hermes et al. (2013) (dotted  red line) and by
Steinfadt et al. (2010) (dashed red line). It seems that our blue edge
and that of Steinfadt et al. (2010) are somewhat cool as compared with
the location  of the  three pulsating ELM  WDs. But, as  mentioned, the
location of  the blue  edge can be  easily accommodated by  tuning the
convective efficiency of the models.

In  Figs.  \ref{figure3}  and  \ref{figure4} we  show the  instability
domains of dipole $g$- and $p$-modes  on the $T_{\rm eff} - \Pi$ plane
corresponding to  our set  of models with  $M_* = 0.17  M_{\odot}$ and
$M_*   =  0.198   M_{\odot}$,  respectively.    Also  shown   are  the
periodicities measured in the three known pulsating ELM WDs. Note that
our theoretical results for the range of periods of unstable $g$-modes
are roughly compatible  with the observations, but there  is not a good
quantitative  agreement  regarding the  effective  temperature of  the
instability domains.  In particular, the short periods suspected to be
due  to $p$-modes  in  SDSS J1112+1117  are roughly  accounted  for by  our
computations  in  the  case  of  models with  $M_*=  0.17  M_{\odot}$,
although  the  effective temperatures  of  the  theoretical
instability domain  are somewhat  lower than the  $T_{\rm eff}$  of the
star.

\section{Summary and future work}
\label{conclusions}

We have presented a brief  description of a thorough theoretical study
on the  seismic properties of  low-mass He-core WDs  based on
fully evolutionary  models representative of  these objects (C\'orsico
et al.  2012).  This  study was aimed  to provide a  theoretical basis
from which  to interpret present  and future observations  of variable
low-mass WDs. As the next step, we are planning to compute a fine grid
(in  stellar mass)  of  fully evolutionary  ELM  WD sequences  derived
consistently from binary  evolution with solar metallicity progenitors
and   different   H   envelope   thicknesses,  to   be   employed   in
asteroseismological  studies. A  good target  for  asteroseismology is
SDSS  J111215+111745  because this  star  exhibits  numerous $g$-  and
(suspected) $p$-modes,  with a robust period  determination.  Also, we
plan to perform extensive stability analysis of these models employing
different efficiencies of convection, and explore mode driving through
the $\epsilon$-mechanism due to H burning.

\acknowledgements 

One  of us  (A.H.C.)   warmly thanks  Prof.   Hiromoto Shibahashi  for
support, that allowed him to attend this conference.

{}

\end{document}